\documentclass[prl,twocolumn,a4paper,amsmath,floatfix,longbibliography]{revtex4-1}
\usepackage[utf8]{inputenc}
\usepackage{graphicx}
\usepackage{amsmath}
\usepackage{color}

\begin{document}

\title{Swim pressure on walls with curves and corners}
\author{Frank Smallenburg}
\author{Hartmut L\"owen}
\affiliation{Institut f\"ur Theoretische Physik II: Weiche Materie, Heinrich-Heine-Universit\"at D\"usseldorf, Universit\"atsstr. 1,
40225 D\"usseldorf, Germany}
% Frank Smallenburg, Ran Ni , Hartmut L\"owen

\begin{abstract}
The concept of swim pressure quantifies the average force exerted by microswimmers on confining walls in non-equilibrium. Here we explore how the swim pressure depends on the wall curvature and on the presence of sharp corners in the wall. For active Brownian particles at high dilution, we present a coherent framework which describes the force and torque on passive particles of arbitrary shape, in the limit of large particles compared to the persistence length of the swimmer trajectories. The resulting forces can be used to derive, for example, the activity-induced depletion interaction between two disks, as well as to optimize the shape of a tracer particle for high swimming velocity. Our predictions are verifiable in experiments on passive obstacles exposed to a bath of bacteria or artificial microswimmers.
\end{abstract}

\maketitle

There is a rapidly growing interest in understanding the physical
principles of microswimmers both for living systems such as bacteria
\cite{Elgeti_Gompper_Winkler_review,Marchetti_review,schwarz2012phase,Aranson,toner2005hydrodynamics}
 and for artificial self-propelled colloidal particles
\cite{Schimansky-Geier,Sen2004,Bechinger}. Swimming of these
active particles occurs at low Reynolds numbers and is a nonequilibrium
phenomenon. Intriguingly, the concepts of equilibrium statistical physics
break down for this new type of "active matter".

In particular, when passive obstacles are exposed to an active suspension, such as a bacterial bath, they experience 
non-thermal fluctuations from the self-propelled particles which result in a fascinating wealth of
new nonequilibrium phenomena ranging from self-starting rotors and cogwheels
\cite{diLeonardo,sokolov2010swimming,Dunkel_PRE_2014} and the spontaneous motion of
microwedges \cite{Kaiser_PRL,kaiser2014motion} to the rectification and sorting of
bacterial motion through asymmetric barriers \cite{Reichhardt,altshuler2013flow,mijalkov2013sorting} and ratchets
\cite{diLeonardo}. This opens  the possibility to power
microengines by an active bath \cite{Kaiser_PRL,nguyen2014emergent} and to steer
micro-robots by activity \cite{de2013three}.
Moreover, more recently, the  swimmer-induced interaction between two parallel
walls \cite{Reichhardt_Casimir,Ran_PRL,Cacciuto}
and two spheres \cite{Egorov_Binder_PRL_2014,angelani2011effective}
has been considered and simulated. This is in a certain
analogy to what is known as depletion \cite{AO} and fluctuation-induced
\cite{Dietrich} forces in equilibrium (or active \cite{schwarz2012phase}) suspensions.

There is a need to understand the principles of these novel
non-equilibrium effects within conceptually simple models which
generalize the concept of equilibrium thermodynamics.
This is most directly done in the context of active Brownian motion describing
self-propelled colloidal spheres. One of the basic
quantities characterizing the interaction between a swimmer bath and an
obstacle is the recently introduced swim pressure \cite{Brady_PRL_2014, yang2014aggregation}.
The swim pressure is defined as the average force per area
 (or per length in two spatial dimensions) exerted by a microswimmer
bath on a planar wall, such that an intensive mechanical quantity can be
defined in nonequilibrium  suspensions  \cite{Cates1,Cates2}.
In the limit of planar walls and high dilution
of spherical swimmers, the swim pressure has been calculated analytically \cite{Brady_PRL_2014, yang2014aggregation}.
Additionally, there has been recent interest in the forces exerted on curved walls by both spherical \cite{mallory2014curvature} and 
elongated \cite{wysocki2015giant} swimmers, as well as the behavior of swimmers in strong confinement \cite{fily2014dynamics,fily2014dynamics2}.

In this article, we consider the swim pressure on curved walls,
including walls with sharp corners. In particular, we study different wall shapes
exposed to dilute suspensions of torque-free, active Brownian particles. 
We obtain a coherent and unified
framework to describe both the motion of obstacles of arbitrary hard shape
in a bacterial bath and the swimmer-induced force between two spheres in close proximity.
Our superposition-like theory applies to any general shape with
length scales much larger than the persistence length of the active particle trajectories. Note that the opposite limit,
where swimmers are trapped in confinements much smaller than their persistence length, was extensively studied by Fily {\it et al} \cite{fily2014dynamics,fily2014dynamics2}.
The theory presented here is semi-empirical, but captures the essential physics similar to the Asakura-Oosawa model for
depletion forces between spheres in a thermal solvent \cite{AO}.

In two spatial dimensions, we confirm that the curvature correction to the swim pressure
scales as $1/R$ for small curvatures \cite{mallory2014curvature} and we propose a fit for this
correction for arbitrary curvature. This is in some formal analogy
to equilibrium thermodynamic for surfaces where the curvature
corrections to the interfacial free energy are of prime interest
and still debated \cite{Roth_PRL_2004,block2010curvature,evans2004nonanalytic}. The corners are treated as a singularity in the shape and studied separately.
The total force is obtained from a  superposition principle. As applications,
we calculate the force acting on some hard particles of V-like and U-like shape, 
and  propose a simple formula for the depletion
force between two disks which generalizes the seminal Asakura-Oosawa
model to an active suspension.
Our predictions are verifiable in experiments on passive obstacles
of different shape exposed to a bath of bacteria or of artificial microswimmers.

In our model, we consider a two-dimensional system consisting of ideal (i.e. non-interacting) self-propelling particles, in contact with one or more flat or curved walls. Apart from the self-propulsion force $f_0$, the particles experience a drag force with friction coefficient $\zeta$, and rotational diffusion of their axis of self-propulsion with a diffusion constant $D_r$. The equations of motions for the position $\mathbf{x}$ and orientation $\phi$ are therefore:
\begin{eqnarray}
\dot{\mathbf{x}} &=& \mathbf{f}_{ext}/\zeta + f_0 / \zeta \mathbf{n} = \mathbf{f}_{ext}/\zeta + v_0 \mathbf{n} \nonumber\\
\dot{\phi}       &=& \sqrt{2 D_r} R(t), \label{eq:eom}
 \end{eqnarray}
where $f_\mathrm{ext}$ is any external force the particles experience from e.g. the confining walls, $v_0 = f_0 / \zeta$ is the self-propulsion velocity, and $R(t)$ represents a delta-correlated stochastic noise term with zero mean and unit standard deviation. Note that the particles do not undergo translational (Brownian) diffusion. 

The only inherent length scale in this model is the persistence length of the particle trajectories $l_p = v_0 / D_r$; the typical distance a particle travels before it rotates significantly away from its original orientation. Additional length scales can be imposed by the shape of the walls in contact with the active bath. 

In several recent studies \cite{Brady_PRL_2014,yang2014aggregation,mallory2014anomalous}, it was shown that for an ideal system of swimmers, the swim pressure on a flat wall in two dimensions is given by:
\begin{equation}
 P_0 = \rho \zeta v_0^2 / 2 D_r, \label{eq:swimpressure}
\end{equation}
where $\rho$ is the bulk density of swimmers. This swim pressure has been shown to behave similar to a thermodynamic state function only if the swimmers experience no external torques and the swim speed is uniform and isotropic throughout the system \cite{Cates1}. 
% If these conditions are not satisfied, the pressure on confining walls is in general not uniform, and dependent on wall-particle interaction.

In order to investigate the effects of curvature on this swim pressure, we calculate the density profile of torque-free swimmers at and near walls that are either curved or contain a corner. To this end, we find the steady-state solution to the Smoluchowski equation for the orientation-dependent density distribution $\rho(\mathbf{x},\phi, t)$ in the system:
\begin{equation}
 \frac{\partial \rho(\mathbf{x},\phi, t)}{\partial t} = - \mathbf{v}(\phi) \cdot \nabla \rho(\mathbf{x},\phi, t)
                                                        + D_r \frac{\partial^2  \rho(\mathbf{x},\phi, t)}{\partial \phi^2}. \label{eq:smolu}
\end{equation}
Here, $\mathbf{v}(\phi) = (v_0 \cos \phi, v_0 \sin \phi)$ is the propulsion velocity of a particle with orientation $\phi$. We find the steady-state solution of this equation numerically on a mesh, for various choices of the hard-wall geometry. In all cases, we use at least $500$ spatial cells, and $64$ discrete orientations for the particles. The cells are adapted to the expected density profile, with smaller cells near the hard boundaries and sharp corners, where rapid changes in local density occur. Note that at the walls, the local density technically diverges, as the swimmers have a finite probability of being located at the wall. To circumvent this issue, the cells directly adjacent to the hard boundaries are chosen particularly small (thickness on the order of $0.001 l_p$). Particles in these cells are considered to be in contact with the wall, and therefore contribute to the pressure experienced by the boundary. We have verified that varying the number and sizes of cells does not significantly affect our results. Note that to avoid non-physical flow of particles at the wall, the cell boundaries of these edge cells must be chosen normal to the wall. Further details are described in the Appendix.

In addition to numerically solving the Smoluchowski equation, we also perform Brownian dynamics simulations of active solutions in contact with hard walls. In these simulations, the point-like active particles do not interact, and move according to the overdamped Brownian equations of motion (Eq. (\ref{eq:eom})). The interaction between the particles and the wall is a purely hard repulsion.

\begin{figure}
\includegraphics[width=0.9\linewidth]{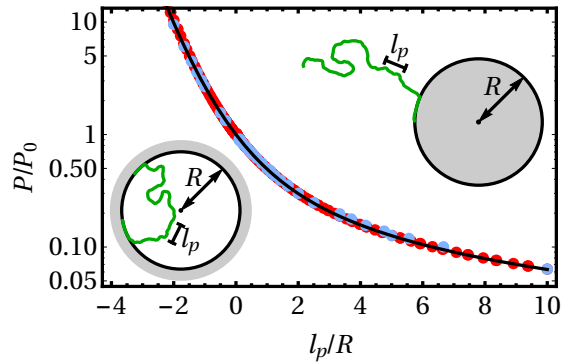}
\caption{Pressure $P$ on walls as a function of the wall radius of curvature $R$. The red points were obtained by numerically solving the Smoluchowski equation, the blue points from simulations, and the black line is the fit from Eq. (\ref{eq:curvefit}). Note that different choices of $l_p$ collapse onto the same curve. The insets show a typical particle trajectory for each geometry.}
\label{fig:curvedwallplot}
\end{figure}

In order to investigate the force exerted on a curved wall by an active solvent, we study both active solutions in spherical confinement, and active solutions in contact with the outside of a circular object, for a range of values of the radius of curvature $R$ of the hard boundary. As expected, in the limit of large $R \gg l_p$, the average pressure on the walls reduces to the swim pressure $P_0$ in Eq. (\ref{eq:swimpressure}). However, when the radius of curvature is finite, particles in contact with the inside of the boundary tend to spend a longer time in contact with the wall than those trapped at the outside. This results in a correction to the swim pressure which increases in magnitude as the radius of curvature decreases. This observation is consistent with recent work by Mallory {\it et al.}, where the authors studied the net motion of curved objects placed in an active bath \cite{mallory2014curvature}. In agreement with their results, we find that the first-order correction to the active pressure is proportional to $1/R$, with the same coefficient for positive and negative radius of curvature. In Fig. \ref{fig:curvedwallplot}, we show the force on both the inside ($R < 0$) and outside ($R>0$) of the wall, as obtained from both simulations and the numerical steady-state solutions of Eq. (\ref{eq:smolu}). We find excellent agreement between both methods. The resulting curve is well fitted by the empirical expression:
\begin{equation}
\frac{P(R)}{P_0} = \alpha e^{-2 l_p/R} + (1 - \alpha) \frac{\log 2}{\log(1 + \exp[l_p/R])},\label{eq:curvefit}
\end{equation}
where $\alpha \simeq 0.0893$ is a single fitting parameter. Note that this expression reduces to $1$ in the limit of flat walls ($R \to \infty$), as expected. We emphasize here that the bulk swim pressure $P_0$ is defined via Eq. \ref{eq:swimpressure} using the density of swimmers measured far away from the wall. In the limit of strong confinement (where $R \ll l_p$), the bulk density tends to zero \cite{fily2014dynamics}, and thus the bulk swim pressure also vanishes. This results in the divergence of Eq. \ref{eq:curvefit} when $l_p / R \to -\infty$. In this limit, nearly all particles are trapped at the wall, and the density distribution on the wall can be described in terms of the local curvature \cite{fily2014dynamics,fily2014dynamics2}.

\begin{figure*}
\begin{tabular}{lll}
 a) & b) & c)\\
\includegraphics[width=0.3\linewidth]{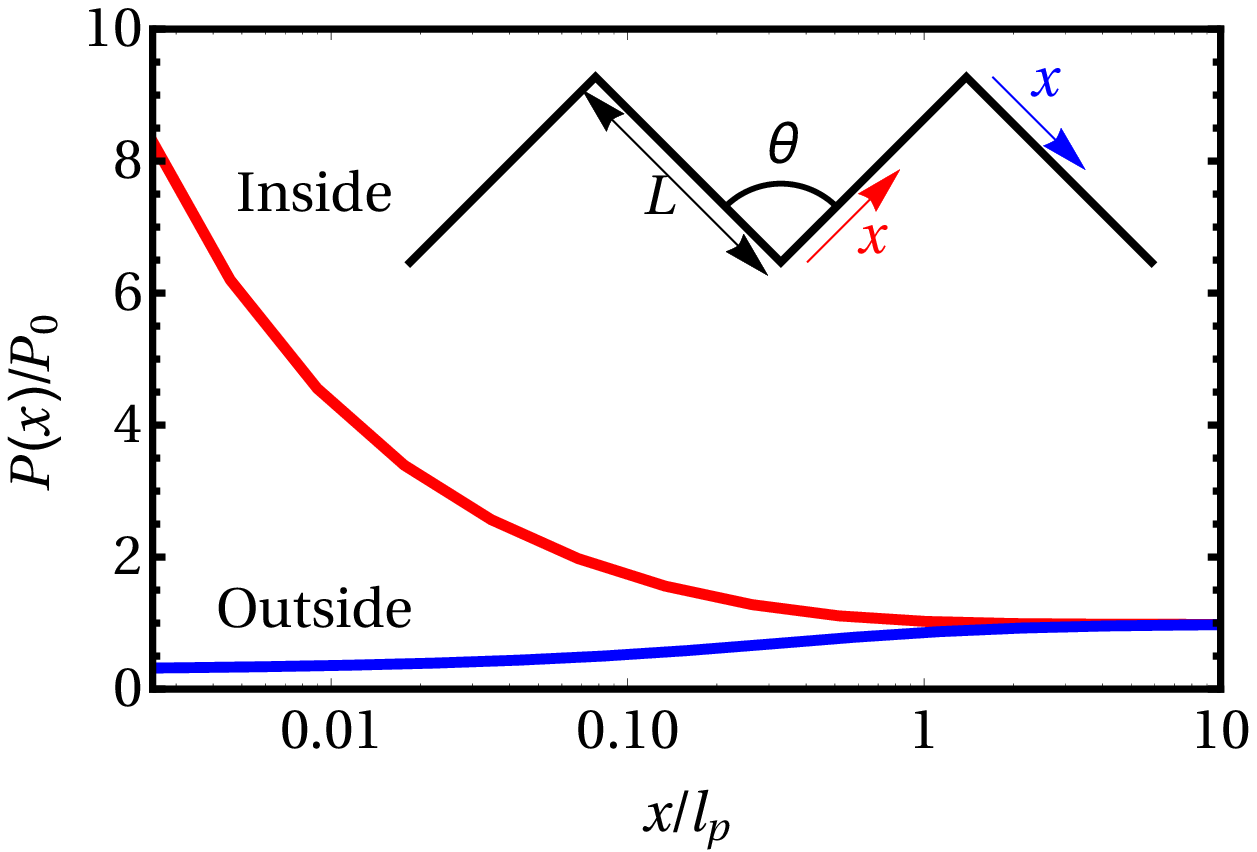} &
\includegraphics[width=0.3\linewidth]{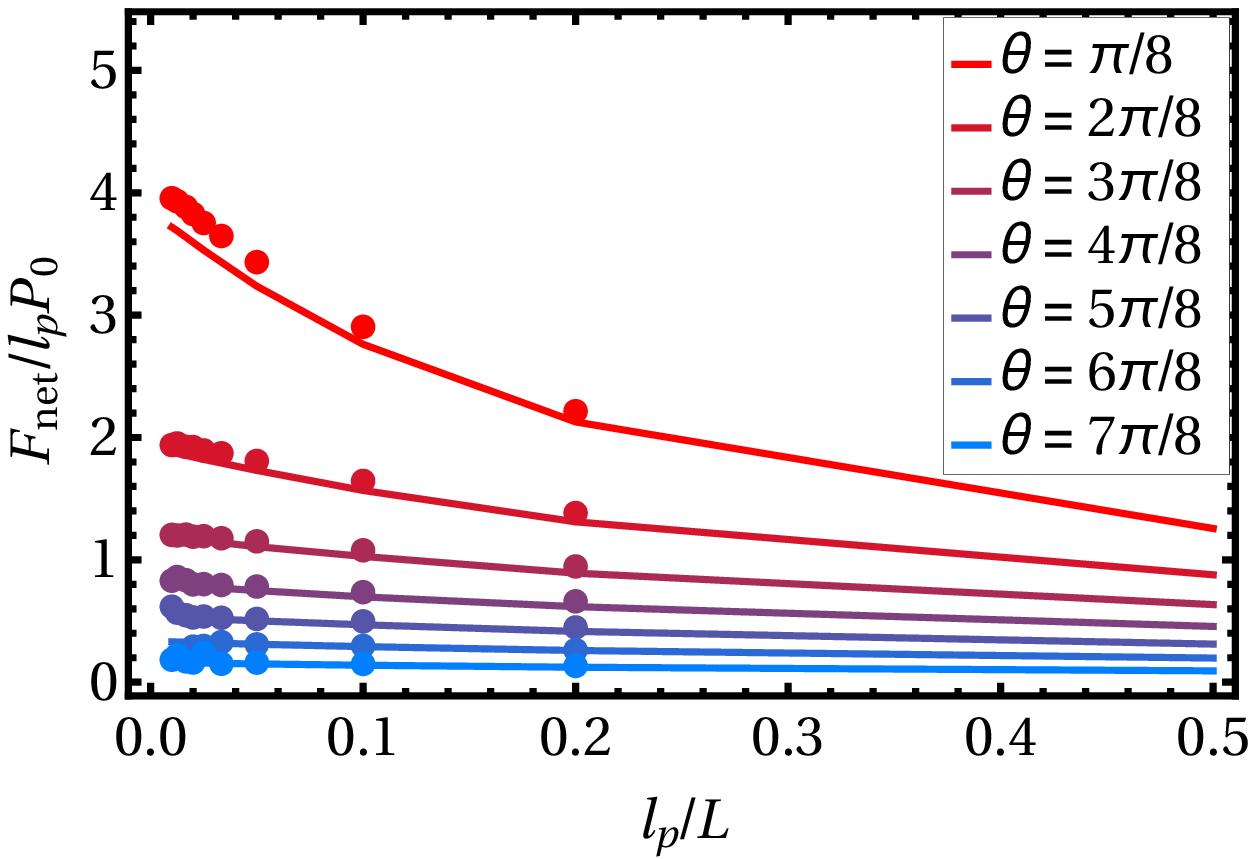} &
\includegraphics[width=0.3\linewidth]{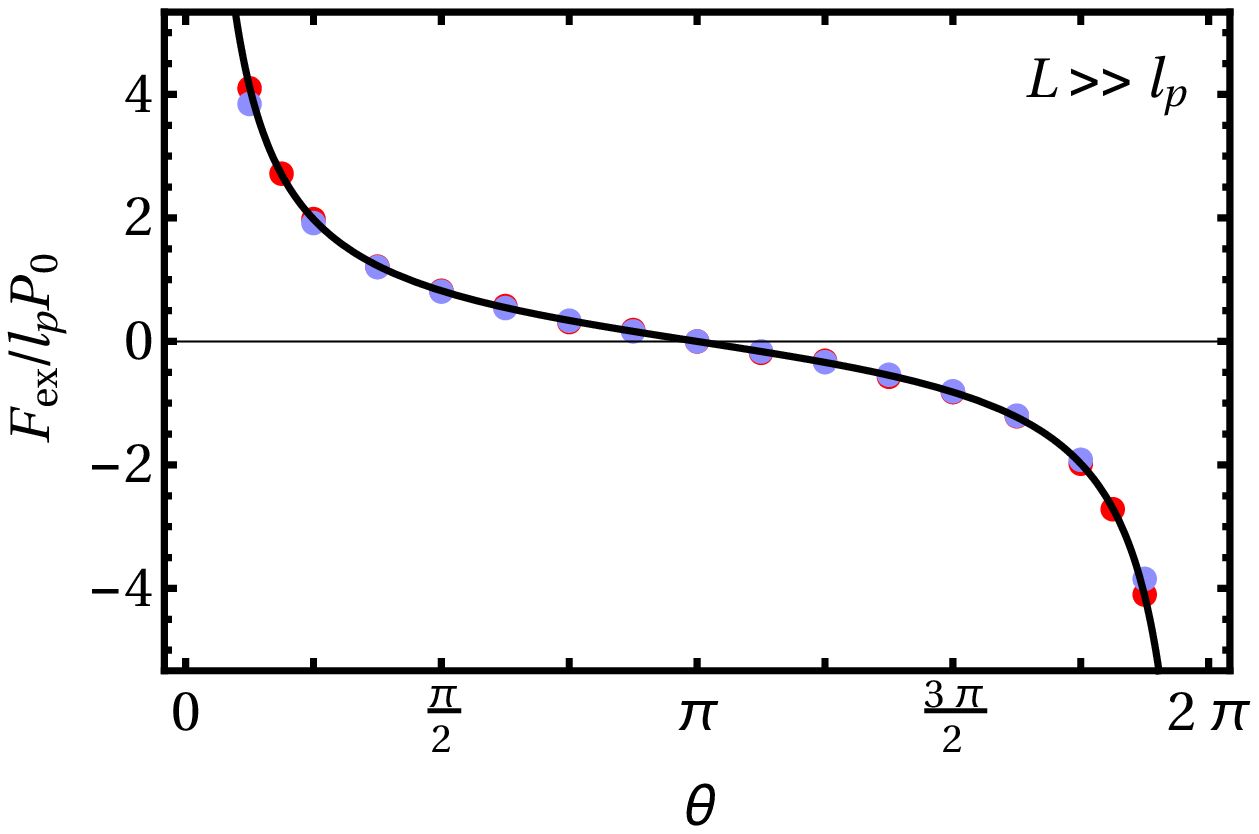}
\end{tabular}
\caption{{\bf a)} Local wall pressure near the inside and outside of a sharp corner of angle $\theta = \pi / 2$, as a function of the distance $x$ to the corner. The corners are part of a wall with a zig-zag geometry, as shown in the inset. Note that far away from the corner, we recover $P=P_0$. {\bf b)} Net force near an inner corner of varying angle for different choices of the ratio of the wall length $L$ and the persistence length $l_p$. The points are the results of Brownian dynamics simulations, the lines are obtained by numerically solving Eq. \ref{eq:smolu}. {\bf c)} Excess force on the wall next to a sharp corner of angle $\theta$, in the limit where the lengths of the (straight) walls adjacent to the corner are much longer than the trajectory persistence length $l_p$. The blue and red points are obtained from the solution to Eq. \ref{eq:smolu} and Brownian dynamics simulations respectively. The black line is the fit in Eq. \ref{eq:forceoncorner}. The force direction is always normal to the wall, and angles $\theta > \pi$ correspond to the outer corner.}
\label{fig:cuspywallplot1}
\end{figure*}

We now consider a sharp corner in a zig-zagging wall, with an opening angle $0<\theta<\pi$, and examine the resulting pressure near both the inner and outer corners. Near the inner corner, increased build-up of particles results in a locally higher pressure (as shown in Fig. \ref{fig:cuspywallplot1}a). In contrast, close to the outer corner the local swim pressure is decreased. We note here that even for relatively long walls ($L/l_p \simeq 20$, there can still be a significant effect of the wall length on the measured force. We note that the total excess force on the wall (as compared to the force resulting from the global swim pressure $P_0$) is zero. This suggests that local irregularities in the geometry of a (globally flat) wall do not interfere with the observation that the pressure of a gas of torque-free swimmers is well-defined \cite{Cates1}. In \ref{fig:cuspywallplot1}b we show for several angles $\theta$ the finite-size scaling for the inner corner for data obtained both from solving the Smoluchowski equation and from simulations.  In Fig. \ref{fig:cuspywallplot1}c we plot the total excess force $F_\mathrm{ex} = F_\mathrm{act} - L P_0$ on the wall, extrapolated to the limit of small $l_p$, as a function of the opening angle. The behavior is well fitted by a function of the form
\begin{equation}
 F_\mathrm{ex}(\theta) = 0.83 P_0 l_p \cot (\theta / 2). \label{eq:forceoncorner}
\end{equation}

The calculated forces can be used to predict the motion of large objects in contact with an active suspension. As an example, we consider a slender object with a continuously varying radius of curvature, for which the shape is parametrized by $\mathbf{w}(s)$. We assume that, compared to the persistence length $l_p$ of the particles, the radius of curvature of the object $R(s)$ is large everywhere, and its rate of change is small. The total force on this object is given by:
\begin{equation}
\mathbf{F} = \int_{s_0}^{s_1} P(R(s)) \hat{\mathbf{n}}(s) \left| \mathbf{w}^\prime(s)\right| \mathrm{d}s \label{eq:forceoncurve},
\end{equation}
where the prime indicates a derivative with respect to $s$, and $\mathbf{n}(s)$ is a vector that is locally normal to the surface of the object.
In the limit of large radius of curvature $R$, the pressure on a curved wall is linear in $1/R$, such that $P(R) = P_0 (1 - A l_p / R)$, with $A \simeq 0.836$. Thus, the net pressure resulting from the forces on both sides of the object is given by $P_\mathrm{net}(R) = 2 A l_p P_0/ R$.  If we assume (without loss of generality), that the curve $\mathbf{w}(s)$ is parametrized by length (i.e.  $\left| \mathbf{w}^\prime(s)\right| = 1$, then $1/R(s) = \frac{\partial \phi(s)}{\partial s}$, where $\phi(s)$ is the local angle  of the curve with the $x$-axis, and Eq. \ref{eq:forceoncurve} reduces to:
% \begin{eqnarray}
% \mathbf{F_\mathrm{act}} &=& 2 A l_p P_0 \int_{s_0}^{s_1} \frac{\hat{\mathbf{n}}(s)}{R(s)}  \mathrm{d}s \\
%            & & 2 A l_p P_0 \int_{s_0}^{s_1} \frac{\partial \phi(s)}{\partial s} \begin{pmatrix}-\sin \phi(s) \\ \cos \phi(s) \end{pmatrix}   \mathrm{d}s \\
%            & & 2 A l_p P_0 \begin{pmatrix} \cos \phi(s_1) - \cos \phi(s_0) \\ \sin \phi(s_1) - \sin \phi(s_0)  \end{pmatrix} \label{eq:forceoncurve2}
% \end{eqnarray}
\begin{equation}
\mathbf{F_\mathrm{act}} = 2 A l_p P_0 \begin{pmatrix} \cos \phi(s_1) - \cos \phi(s_0) \\ \sin \phi(s_1) - \sin \phi(s_0)  \end{pmatrix} \label{eq:forceoncurve2}
\end{equation}
Thus, for curves with a smoothly changing, large radius of curvature, the total force on a curve is only dependent on the angles at the endpoints. 
For objects with sharp corners, the excess force from Fig. \ref{fig:cuspywallplot1}c should be included in $F_\mathrm{act}$ as well.
It is interesting to note that in both cases, the force on an object is independent of the size of the object. Assuming the persistence length of the active solvent is sufficiently small compared to the scale of the object, increasing the length of straight walls does not generate additional net forces. Scaling a curved object, on the other hand, changes the excess pressure on the wall due to the change in radius of curvature. However, in the limit where $R\gg l_p$ this effect is cancelled out by the change in length of the curved region.

In order to illustrate how these results can be used to predict the motion of obstacles in a bath of swimmers, we assume a simple setup where the object and swimmers are suspended in a three-dimensional fluid, but confined to two dimensions via e.g. optical forces. In this case, we can make  use of slender body theory \cite{cox1970motion,batchelor1970slender} to determine the forces exerted on the object by the solvent. For example, the approximate drag force on a curved cylindrical object moving through a solvent moving with velocity $\mathbf{v}$ is given by:
\begin{equation}
 \mathbf{F}_\mathrm{drag} = \frac{4 \pi \mu}{\log (L / a)} \int_0^L  \mathbf{v} \cdot \left( \mathbf{I} - \frac{1}{2} \mathbf{w}'(s) \mathbf{w}'(s) \right)\mathrm{d}s,
\end{equation}
where $L$ is the total length of the curved object, $a$ is the radius of the circular cross-section of the object, $\mu$ is the solvent viscosity, and $\mathbf{I}$ is a unit matrix.
If the object is sufficiently large, we can assume that it will move slowly compared to the velocity of the active particles ($v \ll v_0$), and thus the forces exerted by the active particles will be independent of the object's velocity. In this case, the velocity $\mathbf{v}$ can be calculated simply be setting $\mathbf{F}_\mathrm{drag} = -\mathbf{F}_\mathrm{act}$.  

\begin{figure}
\includegraphics[width=0.8\linewidth]{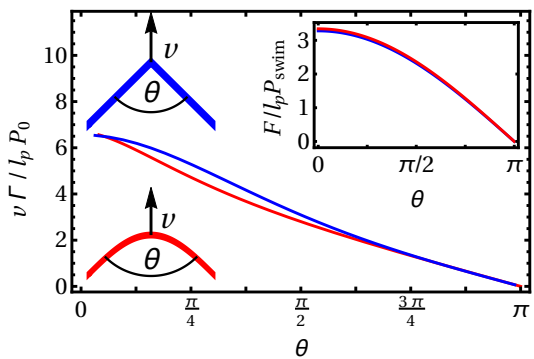}
\caption{Velocity $v$ of a slender U-shape (red) or V-shape (blue) immersed in an active bath. The U-shapes are parametrized by a cosine function of varying amplitude (resulting in a varying angle $\theta$ between the two ends), while the sharp shapes simply consist of two line segments at an angle $\theta$ (see pictures). The object length $L$ is the same in all cases. The prefactor $\Gamma = \log (L/a) / 4 \pi \mu L$, with $\mu$ the viscosity of the solvent. The inset shows that the net force on each shape is approximately the same.}
\label{fig:shapevelocity}
\end{figure}

In Fig. \ref{fig:shapevelocity}, we plot the predicted velocity for both a sharp V-shape and a continuously curved U-shape (parametrized by a cosine function) for a range of opening angles $\theta$ and a fixed total length $L$. These shapes will be propelled forward, in the direction away from the ``open'' end of the shape. When the shape is approximately a straight line ($\theta \simeq \pi$), both shapes move approximately equally fast. However, there are differences between the two velocities at smaller angles. This framework can readily be extended to calculate the net velocity or angular velocity of an arbitrarily shaped slender object, assuming that all of its relevant length scales $L\gg l_p$. Interestingly, these differences purely result from differences in the friction the obstacle experiences from the solvent: as shown in the inset, the net force on each shape is the same for V- and U-shapes within our numerical accuracy. This can also be seen directly from the fact that the prefactor in Eq. \ref{eq:forceoncorner} is approximately equal to $A/2$.

\begin{figure*}
\begin{tabular}{lll}
a)& b)  &c)\\
\includegraphics[width=0.37\linewidth]{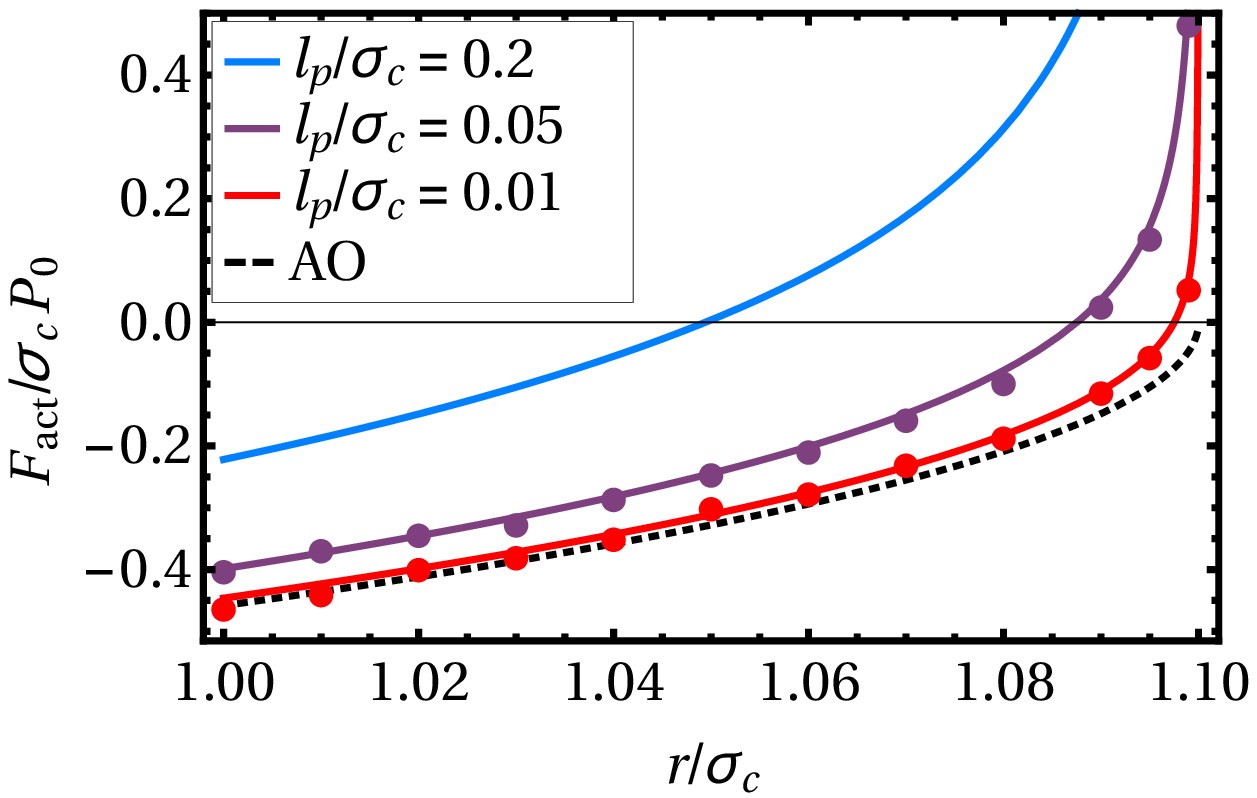} &
\includegraphics[width=0.175\linewidth]{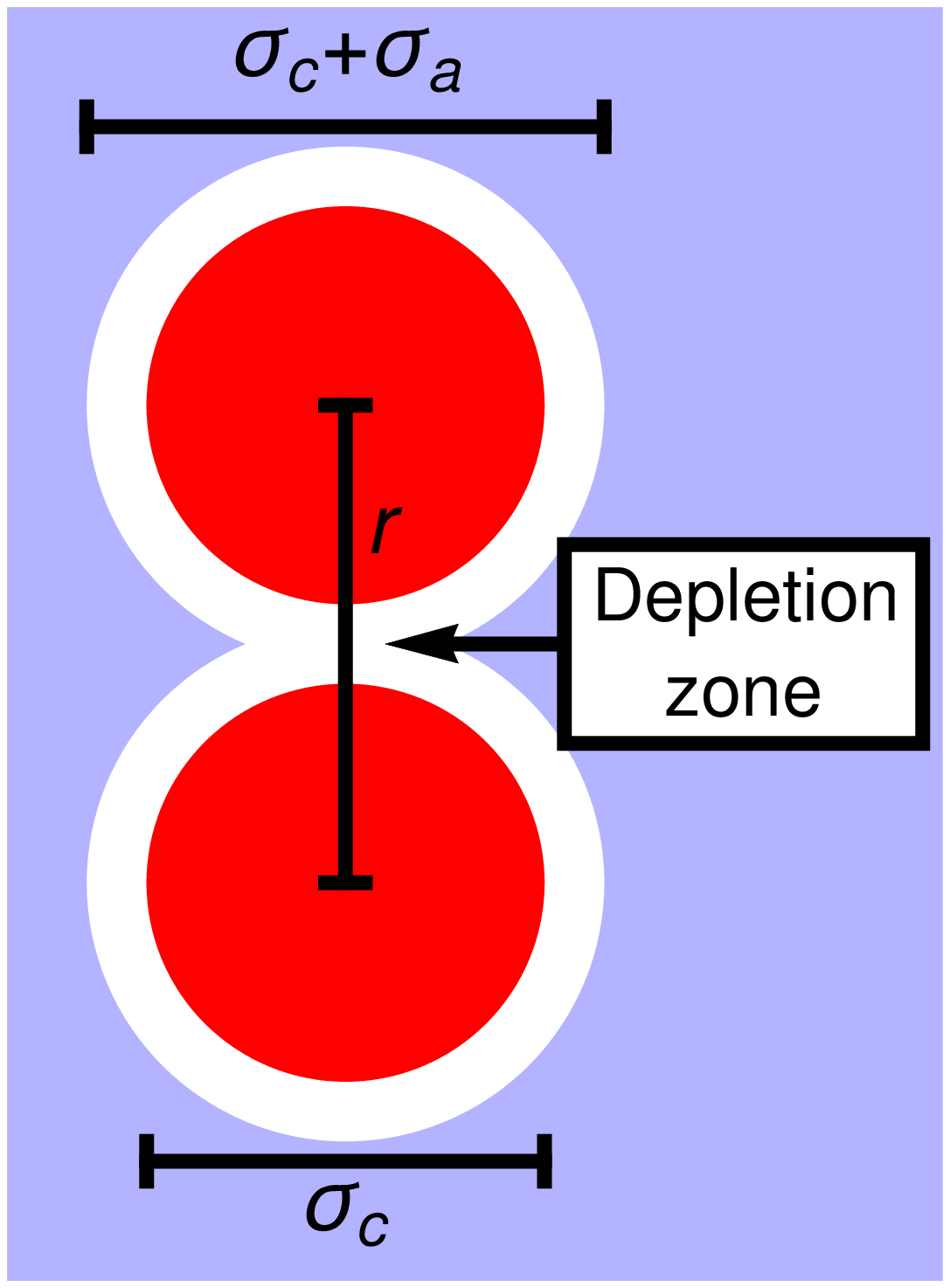} &
\includegraphics[width=0.37\linewidth]{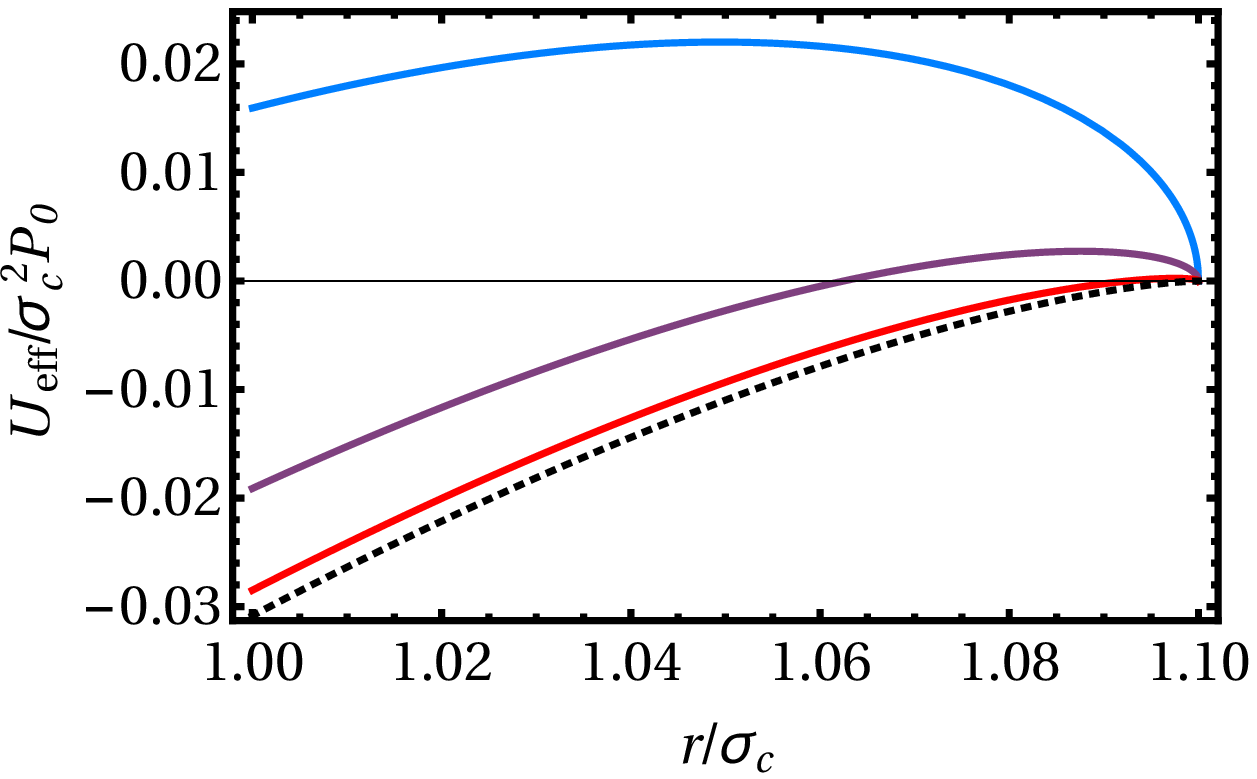} 
\end{tabular}
\caption{{\bf a)} Active depletion force between two disks of diameter $\sigma_c$, immersed in an active bath (particle diameter $\sigma_a = 0.1 \sigma_c$) for various values of the persistence length $l_p$. The lines result from numerical calculations, the points from simulations. The black dashed line is the depletion force in the two-dimensional Asakura-Oosawa model, with the temperature set to the effective temperature $k_B T_\mathrm{eff} = \zeta v_0^2/2 D_r$. Note that a singularity occurs when $r = \sigma_c + \sigma_a$. {\bf b)} Schematic representation of the disks (red), depletion zones (white), and the relevant length scales. {\bf c)} Effective interaction potential $U_\mathrm{eff}$ corresponding to the forces plotted in panel (a). }
\label{fig:depletion}
\end{figure*}

Finally, we use the calculated forces on both curves and corners to predict ``active depletion'' forces: the interaction between two large disks immersed in an active solvent. In this case, we consider two disks of diameter $\sigma_c$, immersed in a bath of particles with diameter $\sigma_a$. The excluded volume for the active particles consists of two overlapping disks of diameter $(\sigma_c + \sigma_a)/2$, with a center-to-center distance $r$. Where the two disks touch, the two sharp corners lead to a repulsive force between the particles, which is counteracted by the active pressure on the rest of the dumbbell shape. The total depletion force is thus given by:
\begin{equation}
 F_\mathrm{depl} = 2 F_\mathrm{ex}(\theta) \sin(\theta) - P(l_p/\sigma_c) \sigma_c \sin (\theta/2),\label{eq:deplforce}
\end{equation}
where $\theta = 2 \arccos (r / \sigma_c)$ is the opening angle at the points where the two disks of excluded volume meet, and $P(l_p/\sigma_c)$ and $F_\mathrm{ex}(\theta)$ are given by Eqs. \ref{eq:curvefit} and \ref{eq:forceoncorner}, respectively.

In Fig. \ref{fig:depletion}a, we plot the active depletion force as a function of distance for several choices of the persistence length. In the limit of small $l_p$, the pressure on the curved walls dominates the forces on the corners, and we recover the two-dimensional Asakura-Oosawa depletion force, with an effective temperature $k_B T_\mathrm{eff} = \zeta v_0^2/2 D_r$ (such that $P_0$ is equivalent to an ideal gas pressure). We obtain good agreement between simulations and the prediction of Eq. \ref{eq:deplforce} for small values of $l_p / \sigma_c$. We can obtain an effective interaction potential, $U_\mathrm{eff}(r)$ by integrating the obtained forces:
\begin{equation}
 U_\mathrm{eff}(r) = -\int_{\sigma_c+\sigma_a}^{r} \mathrm{d}r' F_\mathrm{act}(r').
\end{equation}
where we have chosen the interaction potential to be zero at $r = \sigma_c+\sigma_a$. We plot the effective potentials in Fig. \ref{fig:depletion}c. Note that despite the divergence in the force (Fig. \ref{fig:depletion}a), the interaction energy remains finite, indicating that the colloids can in principle approach each other.

Note that for distances larger than $r = \sigma_c+\sigma_a$, the active particles can suddenly pass between the two disks, leading to a discontinuous change in the force. Although we expect the depletion force to decay rapidly to zero beyond this distance, we have not closely investigated this regime, as it inherently involves geometrical features with a length scale smaller than $l_p$. However, simulations indeed show a steep drop to zero in the force at slightly larger distances. Moreover, in a recent study on active depletion forces \cite{Cacciuto}, simulations for relatively weak active propulsion (i.e. small $l_p$) indeed revealed a peak in the depletion force near  $r = \sigma_c+\sigma_a$, and a decay to zero at larger distances. Note that more complex effects, such as geometric shielding, can occur when the persistence length of the particles is large compared to the separation of the two obstacles. In particular, simulations of swimmers confined between two walls \cite{Reichhardt_Casimir} showed oscillatory behavior of the interaction force in this regime.

% \section{Conclusions}

In conclusion, we have generalized the concept of swim pressure to non-flat walls,
and calculated the curvature- and corner-induced corrections to the bulk swim pressure both numerically and via particle-resolved simulations.
This approach yields a coherent framework to obtain the forces and torques on passive particle of arbitrary shape and to derive a simple
approximation for the activity-induced depletion interaction. 

Our approach can in principle be generalized to three spatial dimensions as well as to incorporate more details of the hydrodynamic swimmer-wall interactions.
It would further be interesting to study the motion of passive carriers in a density gradient of active particles. Moreover, an important extension
of this work would be to consider an active background outside the dilute regime, where the interactions between the active particles can lead to clustering, ordering,
and phase separation \cite{buttinoni2013dynamical,fily2014freezing,palacci2013living,mognetti2013living,zottl2014hydrodynamics,huepe2004intermittency}.

\section{Appendix: Computational Details}

In order to numerically obtain the density distribution of active particles near an obstacle, we find the steady-state solution of the the Smoluchowski equation for the orientation-dependent density distribution $\rho(\mathbf{x},\phi, t)$:
\begin{equation}
 \frac{\partial \rho(\mathbf{x},\phi, t)}{\partial t} = - \mathbf{v}(\phi) \cdot \nabla \rho(\mathbf{x},\phi, t)
                                                        + D_r \frac{\partial^2  \rho(\mathbf{x},\phi, t)}{\partial \phi^2} = 0. \label{eq:smoluSI}
\end{equation}
Here, $\mathbf{x}$ indicates the particle position, $\phi$ denotes the particle orientation, and $\mathbf{v}(\phi) = v_0 (\cos \phi, \sin \phi)$ is the self-propulsion velocity of a particle with orientation $\phi$, where $v_0$ is the swim speed. $D_r$ is the orientational diffusion coefficient. 
To find the steady-state solution to this equation, we divide the system area up into $n_{xy}$ cells, and divide the set of possible orientations into $n_\phi$ equal intervals. We assume that the cells are sufficiently small, such that in each cell $\rho(\mathbf{x},\phi)$ is approximately constant over the area and orientation range of the cell, and calculate the total flow of particles into and out of each cell. For each cell, this particle current can be written as a linear combination of the densities of the cell and its neighbors. Setting the rate of density change for each cell to zero yields a complete set of linear equations for the density $\rho_{i,k}$ for each cell $i$ and orientation $k$, of the form:
\begin{equation}
 \frac{\partial \rho_{i,k}}{\partial t} = v_0 \sum_{j=1}^{n_{xy}} T_{ij}(k) \rho_j(\phi_k) + D_r \sum_{l=1}^{n_\phi} R_{kl}(i) \rho_i(\phi_l)  = 0, \label{eq:steadystatematrix}
\end{equation}
where the translation matrix $T_{ij}(k)$ is the $n_{xy} \times n_{xy}$ matrix representing the flow of particles with orientation $k$ between cells due to their active motion. Similarly, the rotation matrix $R_{kl}(i)$ represents the rotational diffusion in each cell $i$. In order to obtain the elements of the translation matrix, we determine the particle current $J_{c,e}$ out of cell $c$ through edge $e$ with edge length $l_e$ and norm $\mathbf{n}_e$ using an upwind scheme, yielding:
\begin{equation}
 J_{c,e} = \left\lbrace
\begin{array}{ll}
\mathbf{v}(\phi_k)\cdot \mathbf{n}_e l_e \rho_{c,k} & \text{if }\mathbf{v}(\phi_k)\cdot \mathbf{n}_e \ge 0\\
\mathbf{v}(\phi_k)\cdot \mathbf{n}_e l_e \rho_{\mathrm{neighbor},k} & \text{otherwise} \\
\end{array}
\right.,
\end{equation}
where $\phi_k = 2 k \pi / n_\phi$, and $\rho_{\mathrm{neighbor},k}$ indicates the density of the cell neighboring cell $c$ through edge $e$. 
Combining all particle currents yields the translation matrix:
\begin{eqnarray}
 \left(\frac{\partial \rho_{i,k}}{\partial t}\right)_\mathrm{trans} &=& \sum_{e \in \mathrm{edges}(i)} \frac{F_{i,e}}{A_i}, \\
                                                                    &=& v_0 \sum_{j \in \mathrm{cells}} T_{ij}(k) \rho_{j,k} 
\end{eqnarray}
where $A_i$ is the surface area of cell $i$.
The rotation matrix $R_{kl}(i)$ is obtained by numerically approximating the second-order derivative with respect to $\phi$ in Eq. \ref{eq:smoluSI}:
\begin{eqnarray}
  \left(\frac{\partial \rho_{i,k}}{\partial t}\right)_\mathrm{rot} &=& D_r \frac{ \left[\rho_{i,k+1} - \rho_{i,k}\right] - \left[\rho_{i,k} - \rho_{i,k-1}\right]}{ (2\pi / n_\phi)^2 }\\
 &=& D_r \sum_l R_{kl}(i) \rho_{i,l}
\end{eqnarray}

Using Eq. (\ref{eq:steadystatematrix}), the translation and rotation matrices define a full set of linear equations for the density distribution in the system, which can be solved numerically. The hard walls in the system are represented by zero-flux boundary conditions on parts of the mesh edges. For the other mesh boundaries, we make use of either reflecting or shifted boundary conditions, depending on the symmetry of the geometry. For the case of curved walls, we make use of the rotational symmetry that the density profile exhibits both on the inside and the outside of a circular wall. In this case, only one row of cells is needed along the radial direction, and translation along the azimuthal direction can be considered as a change in orientation instead of a change in position. In the case of walls with sharp corners, we implement a wall with a zigzagging geometry, where we make use of mirror symmetry in order to reduce the size of the mesh. Additionally, to reduce numerical error, we make use of the knowledge that the total wall pressure should equal the bulk swim pressure (Eq. \ref{eq:swimpressure}).

\acknowledgments{
We would like to thank R. Ni, D. Takagi, B. ten Hagen, R. Wittkowski, and T. Speck for fruitful discussions.
We acknowledge funding from the Alexander von Humboldt foundation and from the DFG, within the Science Priority Programme SPP 1726 on microswimmers.
}

% \bibliographystyle{prsty}
% \bibliography{active}

\end{document}